\documentclass{pasj00}

\begin{document}

\SetRunningHead{Suzaku Studies of the LMXB Aql X-1}{Sakurai et al.}

\title{Suzaku Studies of Luminosity-Dependent Changes in the Low-Mass X-ray Binary Aquila X-1}


 \author{%
   Soki \textsc{Sakurai}\altaffilmark{1}
   Shunsuke \textsc{Torii}\altaffilmark{1}
   Hirofumi \textsc{Noda}\altaffilmark{1}
   Zhongli \textsc{Zhang}\altaffilmark{1}
   Ko \textsc{Ono}\altaffilmark{1}
   Kazuhiro \textsc{Nakazawa}\altaffilmark{1}
   Kazuo \textsc{Makishima}\altaffilmark{1,2}
   Hiromitsu \textsc{Takahashi}\altaffilmark{3}
   Shin'ya \textsc{Yamada}\altaffilmark{4}
   and
   Masaru \textsc{Matsuoka}\altaffilmark{2}
   }
 \altaffiltext{1}{Department of Physics, University of Tokyo 7-3-1, Hongo, Bonkyo-ward, Tokyo, 113-0033}
 \email{sakurai@juno.phys.s.u-tokyo.ac.jp}
 \altaffiltext{2}{MAXI Team, Institute of Physical and Chemical Research, 2-1 Hirosawa, Wako, Saitama, 351-0198}
 \altaffiltext{3}{High Energy Astrophysics Group, Department of Physical Sciences
Hiroshima University
1-3-1 Kagamiyama, Higashi-Hiroshima, Hiroshima, 739-8526}
 \altaffiltext{4}{High-Energy Astrophysics Laboratory , Institute of Physical and Chemical Research, 2-1 Hirosawa, Wako, Saitama, 351-0198}



%


\maketitle

\begin{abstract}
The neutron-star Low-Mass X-ray Binary Aquila X-1 
was observed by Suzaku for seven times, from 2007 September 28 to October 30.
The observations successfully traced an outburst decay in which the source luminosity
decreased almost monotonically from $\sim 10^{37}$ erg s$^{-1}$ to $\sim 10^{34}$ erg s$^{-1}$, by $\sim 3$ orders of magnitude.
To investigate luminosity-dependent changes in the accretion geometry,
five of the seven data sets with a typical exposure of $\sim 18$ ks each were analyzed; 
the other two were utilized in a previous work \citep{Sakurai2012}.
The source was detected up to 100 keV in the 2nd to the 4th observations, 
to 40 keV in the 5th, and to 10 keV on the last two occasions.
All spectra were reproduced successfully by Comptonized blackbody model
with relatively high ($\gtrsim 2.0$) optical depths, plus an additional softer optically-thick component.
The faintest three spectra were reproduced alternatively by a single Comptonized blackbody model
with a relatively low ($\lesssim 0.8$) optical depth.
The estimated radius of the blackbody emission, including seed photons for the Comptonization, 
was $10 \pm 2$ km at a 0.8--100 keV luminosity of $2.4\times 10^{36}$ erg s$^{-1}$
(the 2nd to the 4th observations).
In contrast, it decreased to $7 \pm 1$ km and further to $3 \pm 1$ km, at a luminosity of $(4.8-5.2)\times 10^{35}$ erg s$^{-1}$
(the 5th observation) and $\sim 2\times 10^{34}$ erg s$^{-1}$ (the 6th and 7th), respectively,
regardless of the above model ambiguity.
This can be taken as evidence for the emergence of a weak magnetosphere of the neutron star.
\end{abstract}

\section{Introduction}
	A neutron-star Low-Mass X-ray Binary (NS-LMXB) consists of a Roche-lobe-filling low-mass star and a mass-accreting neutron star 
with a weak magnetic field of $B < 10^9$ G.
These X-ray sources generally have two spectral states, namely the soft and hard states 
(e.g. Mitsuda et al. 1984; Mitsuda et al. 1989; White \& Mason 1985; White et al. 1988).
In the soft state, the X-ray spectrum is relatively soft, and comprises two major optically-thick emission components.
The softer of them is a disk blackbody with a temperature of $0.5-2$ keV, and the harder one is a blackbody of 
temperature $\sim 2$ keV, often weakly Comptonized as evidenced by data excess towards higher energies \citep{Barret2001,Iaria2005, Lin2007}.
In addition, a harder emission component, possibly of non-thermal nature, may underlie the optically-thick spectrum \citep{Lin2007}.

In the hard state, the spectrum has a hard power-law-like shape that also consists of two emission components.
a soft thermal component and a hard one with strong Comptonization.
There have been a degeneracy between two modelings of the emission.
One invokes a soft disk blackbody plus a Comptonized blackbody, while the other 
a soft blackbody plus a hard Comptonized disk blackbody \citep{Gierlinski2002,Barret2003,Lin2007}.
\citet{Sakurai2012}, hereafter Paper 1, showed that the former modeling is preferred by broad-band Suzaku data
of the NS-LMXB Aquila X-1 (hereafter Aql X-1): the blackbody emission from the neutron-star surface is
strongly Comptonized by a hot spherical corona, 
resulting in a hard continuum approximated by a power-law photon index of $\Gamma\sim 2$.

\begin{table*}[htb]
\caption{Suzaku observations of Aql X-1 in 2007.}
\centering
\begin{minipage}{11cm}
\begin{tabular}{ccccccc}\hline
tag&Date\footnotemark[$*$]  & ObsID & Exp\footnotemark[$\dagger$] & \multicolumn{3}{c}{Count rate\footnotemark[$\ddagger$] (counts/s)}
\\\cline{5-7} 
&&& (ks) & XIS\footnotemark[$\S$]  & HXD-PIN\footnotemark[$\|$] & HXD-GSO\footnotemark[$\#$] \\\hline
1&9/28 & 402053010 & 13.8 & 275.4$\pm$0.4\footnotemark[$**$]  & 1.12$\pm$0.01 & -\\
2&10/03 &402053020&15.1  & 28.12$\pm$0.04 & 0.74$\pm$0.01 & $0.26\pm 0.04$\\
3&10/09 &402053030&19.7  & 33.18$\pm$0.04 & 0.89$\pm$0.01 & $0.19\pm 0.03$\\
4&10/15 &402053040&17.9  & 25.85$\pm$0.04 & 0.73$\pm$0.01 &$0.14\pm 0.03$ \\
5&10/19 &402053050&17.9  &  5.67$\pm$0.02 & 0.13$\pm$0.01 & -\\
6&10/24 &402053060&21.4  & 0.221$\pm$0.003 & 0.041$\pm$0.005 & -\\
7&10/30 &402053070&17.5  & 0.172$\pm$0.003 & 0.056$\pm$0.006 & -\\\hline
\end{tabular}
\label{table_obsproperty}
\footnotetext[$*$]{Date in 2007 as Month/Day.}
\footnotetext[$\dagger$]{Net exposure per XIS sensor.}
\footnotetext[$\ddagger$]{In ct s$^{-1}$ after subtracting the background.  The quoted errors are statistical 1$\sigma$ limits.}
\footnotetext[$\S$]{In the 0.8--10 keV range with the two FI sensors (XIS0, XIS3) summed.}
\footnotetext[$\|$]{Referring to a 12--50 keV range, including the CXB contribution of $\sim 0.02$ ct s$^{-1}$.}
\footnotetext[$\#$]{Referring to a 50--100 keV range. The CXB contribution is negligible because it is less than $0.1$\% of the total GSO background.}
\footnotetext[$**$]{Corrected for pile-up effects; a central region within $1'$ of the XIS image was discarded.}
\end{minipage}
\end{table*}%
    
	In the present research which follows Paper 1, we further investigate spectra of Aql X-1.
It is a transient recurring with typical intervals of $\sim 1$ yr,
and its distance is estimated by \citet{Jonker2004} as 4.4--5.9 kpc.
In the decay phase of an outburst which occurred in 2007 September--October, it was observed with 
the X-ray observatory Suzaku \citep{Mitsuda2007} for seven times,
wherein the source luminosity decreased from $\sim 10^{37}$ erg s$^{-1}$ to $\sim 10^{34}$ erg s$^{-1}$.
Among these, the first four were analyzed by \citet{Raichur2011}, and the first and third ones in Paper 1.
These results have shown that the source resided in the typical soft state in the first observation,
and in the hard state in the second to forth ones.
In the present paper, we analyze five out of the seven data sets, excluding the 1st and the 3rd ones which were
already analyzed in Paper 1.
By thus combining all the seven data sets, we would like to study how the accretion-flow geometry changed as the luminosity varied by $\sim 3$ orders of magnitude.

\section{Observation and Data Processing}
The log of the seven Suzaku observations are given in table \ref{table_obsproperty}, 
which reproduces and updates (by adding the last column) table 1 of Paper 1.
Thus, except for a possible re-brightening in Observation 3 (hereafter Obs$.$ 3), the source intensity decreased monotonically.
However, the decline rate from one observation to the next, which are separated by $4-6$ days, varied very much.
In particular, the XIS intensity decreased rapidly from Obs$.$ 5 to Obs$.$ 6.


\subsection{Data Processing}\label{ss:dp}

\subsubsection{XIS data processing}\label{ss:dp:xis}
\begin{figure*}[htb]
	\begin{center}
		\FigureFile(80mm,80mm){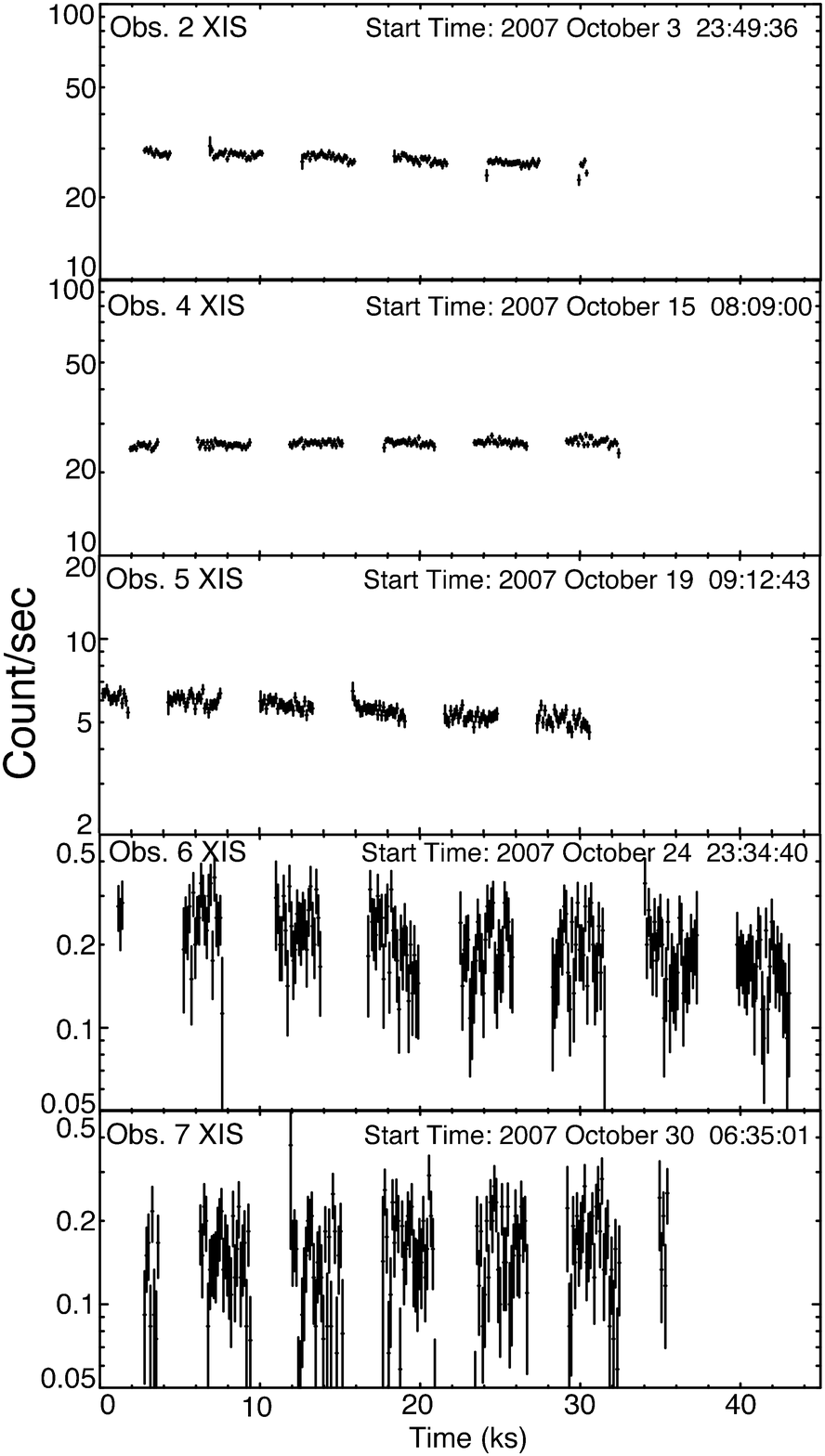}
		\FigureFile(80mm,80mm){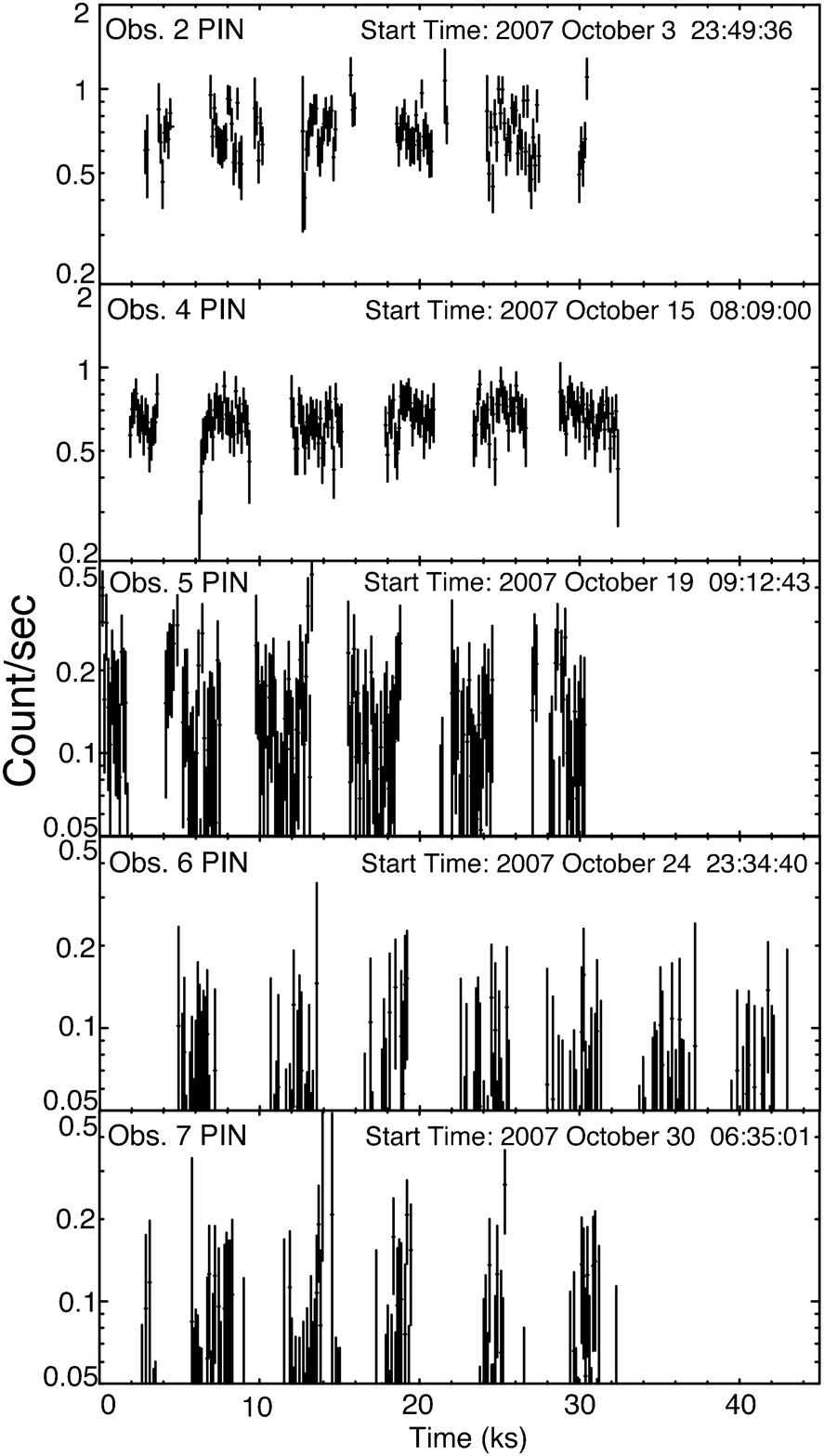}
	\end{center}
  \caption{ Background-subtracted light curves from XIS FI (left-side panels, 0.8--10 keV)
  and HXD-PIN (right side, 12--50 keV), presented with a binning of 120 s.
  The time origin is given in each panel.
  The HXD-PIN results include the CXB but not the NXB. 
  Ordinate is logarithmic, spanning an order of magnitude.
  The HXD signal detection is insignificant in Obs$.$ 6 and Obs$.$ 7.
}\label{fig_lc}
\end{figure*}

	The	 XIS data of the five observations were all processed in the same manner as the 3rd one in Paper 1.
Their XIS light curves are shown in figure \ref{fig_lc}.
None of them shows burst-like events, with low (typically $<20\%$) variability.
Therefore, like in Paper 1, we accumulated the entire events of the two FI sensors (XIS0+XIS3) 
from each observation into a time-averaged 0.8--10 keV spectrum.
We again excluded the two energy ranges for Obs$.$ 2-5, 
1.7--1.9 keV and 2.2--2.4 keV, to avoid the instrumental silicon K-edge and gold M-edge, respectively,
where calibration uncertainties are large.
In Obs$.$ 6 and Obs$.$ 7, when the count rates was low, we used in contrast the full $0.8-10$ keV range, 
because the statistical error dominates over such instrumental effects.

\subsubsection{HXD data processing}\label{ss:dp:hxd}
	In all observations, the HXD data were screened by the same criteria as in Paper 1.
In short, the Non X-ray Background (NXB) was subtracted from both the light curves and spectra,
while contribution from the Cosmic X-ray Background (CXB) to the spectra was separately modeled as
\begin{equation}
\textmd{CXB}(E) = 9.41\times 10^{-3}  \left(\frac{E}{1\textmd{keV}}\right)^{\hspace{-0.3em}-1.29}\hspace{-1.4em}\exp\left(-\frac{E}{40\textmd{keV}}\right)  ,
\label{eq:cxb}
\end{equation}
where the unit is photons cm$^{-2}$s$^{-1}$keV$^{-1}$FOV$^{-1}$.

In Obs$.$ 2 and Obs$.$ 4, the HXD-PIN and HXD-GSO signals were successfully detected 
over 12--50 keV and 50--100 keV, respectively.
The obtained HXD-PIN light curves are shown in figure \ref{fig_lc} (right column), 
while the HXD-PIN and HXD-GSO source count rates (after removing the background)
are given in table \ref{table_obsproperty}, which partially reproduce the information already given in Paper 1.
In Obs$.$ 5, the HXD-PIN signals were detected over 12--50 keV and yielded a light curve in figure \ref{fig_lc}, while the HXD-GSO data were consistent with background
within statistical errors.
The HXD-PIN data in the last two observations were not used because their count rates are consistent 
with the expected CXB intensity (see figure \ref{fig_lc}), when systematic errors of the NXB subtraction is taken into account.
 The HXD-GSO signals were not detected, either.


\section{Spectral analysis}\label{s:spectral_analysis}
\subsection{Overview}\label{ss:sa:overview}
\begin{figure}[htb]
	\begin{center}
		\FigureFile(80mm,80mm){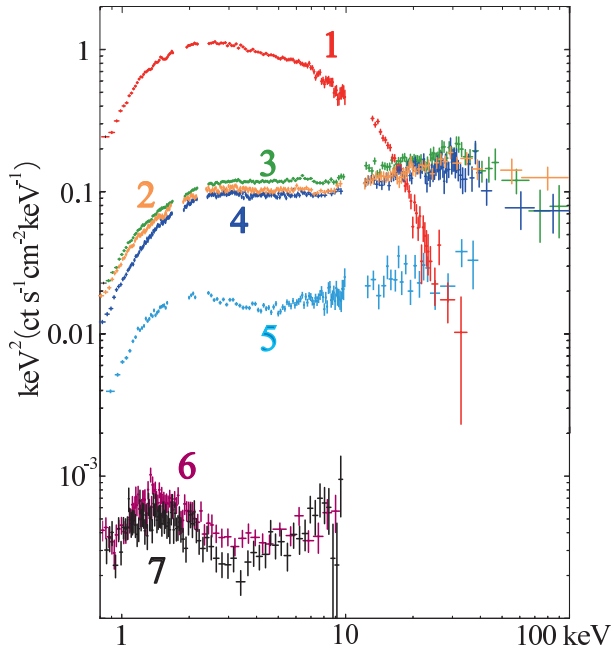}
	\end{center}
  \caption{Background-subtracted and response-removed $\nu F \nu$ spectra of Aql X-1 taken with Suzaku in Obs$.$ 1 (red), Obs$.$
  2 (orange), Obs$.$ 3 (green), Obs$.$ 4 (dark blue), Obs$.$ 5 (light blue), Obs$.$ 6 (purple), and Obs$.$ 7 (black).
  The CXB contribution has been subtracted from the HXD-PIN data, which covers the 12--50 keV energy range.
  }
  \label{fig_allobs}
\end{figure}

Figure \ref{fig_allobs} shows the seven spectra which were obtained in this way.
It is clear that the spectral shape changed drastically as the source luminosity decreased.
In particular, the spectrum of Obs$.$ 1 is much softer than the others, with a steep decline in energies above $\sim 10$ keV.
The spectra from Obs$.$ 2, Obs$.$ 3, and Obs$.$ 4 are very similar with an approximately power-law-like shape
spanning a very broad energy range with an approximate
photon index of $\Gamma\sim 2$, possibly with a cutoff at $\gtrsim 30$ keV.
That of Obs$.$ 5 is also power-law-like, similar to those in Obs$.$ 2 to 4,
but is more concave around 4 keV.
In the last two observations, the spectrum commonly has a charasteristic shape 
possibly consisting of a soft optically-thick emission plus a very hard power-law component, 
in agreement with previous studies in the quiescent state of this source \citep{Rutledge2002,Campana_Stella2003,Cackett2011}.

Below, the five spectra are classified into three groups by their spectral shapes and fluxes; 
the first group consists of Obs$.$ 2 and Obs$.$ 4, which are very similar to Obs$.$ 3 analyzed in Paper 1.
The second group is Obs$.$ 5, while the last group comprises Obs$.$ 6 and Obs$.$ 7.
In the following model fits, the effects of interstellar absorption are represented by a model {\tt wabs} in {\tt xspec}.
An updated absorption model, tbabs with an abundance table of {\tt wilm} \citep{Wilms2000}, gives essentially the same results,
except systematically (by $\sim 20\%$) higher value for the equivalent hydrogen column density $N_{\rm \scriptsize{H}}$.
For the sake of consistency with Paper 1, we retain the use of {\tt wabs}.
When the data suggest the presence of an optically-thick disk emission, we describe it with a {\tt diskBB} model,
and estimate the innermost disk radius $R_{\rm in}$ from the best-fit model normalization assuming that the
obtained radius parameter is in
fact equals to  $\xi^{-1} \kappa^{-2} R_{\rm in} (10\; {\rm kpc}/d)^2\sqrt{\cos i}$.
Here, $\xi=0.412$ is a correction factor for the inner boundary condition \citep{Kubota1998,Makishima2000}, 
$\kappa =1.7$ is the standard color hardening factor \citep{Shimura_Takahara1995}, $d$ is the source distance, and $i$ is the disk inclination for which we assumed a value of $45^\circ$ since it is not known.
These assumptions are just the same as those employed in Paper 1.


\subsection{Obs$.$ 2 and Obs$.$ 4}\label{ss_obs2and4}
	As seen in figure \ref{fig_allobs}, the spectra from these two data sets are both very similar 
in shape and intensity to that of Obs$.$ 3 analyzed in \citet{Raichur2011} and Paper 1.
In Paper 1, the spectrum of Obs$.$ 3 was reproduced successfully over the 0.8--100 keV range, 
by a weak disk blackbody, and a Comptonized blackbody with an electron temperature of
$T_\textmd{\scriptsize{e}}\sim 30$ keV and an optical depth of $\tau \sim 3$.
These parameters are reproduced in table \ref{table_2ndand4th}.

This model for Obs$.$ 3 in fact has the same composition as that for the soft state (as already discussed in Paper 1).
However, the value of $T_\textmd{\scriptsize{e}} \sim 30$ keV characterizing 
its Comptonized blackbody component for the hard state (Obs$.$ 2) 
is an order of magnitude higher than that for the soft state (Obs$.$ 1 in Paper 1),
as evidenced by the very clear spectral difference between Obs$.$ 1 and Obs$.$ $2-4$.
This difference agrees with the general understanding \citep{Gierlinski2002,Falanga2006,Fiocchi2007}
that the Comptonizaion effects become much stronger in the hard state than in the soft state.

\begin{figure*}[htb]
	\begin{center}
		\FigureFile(80mm,80mm){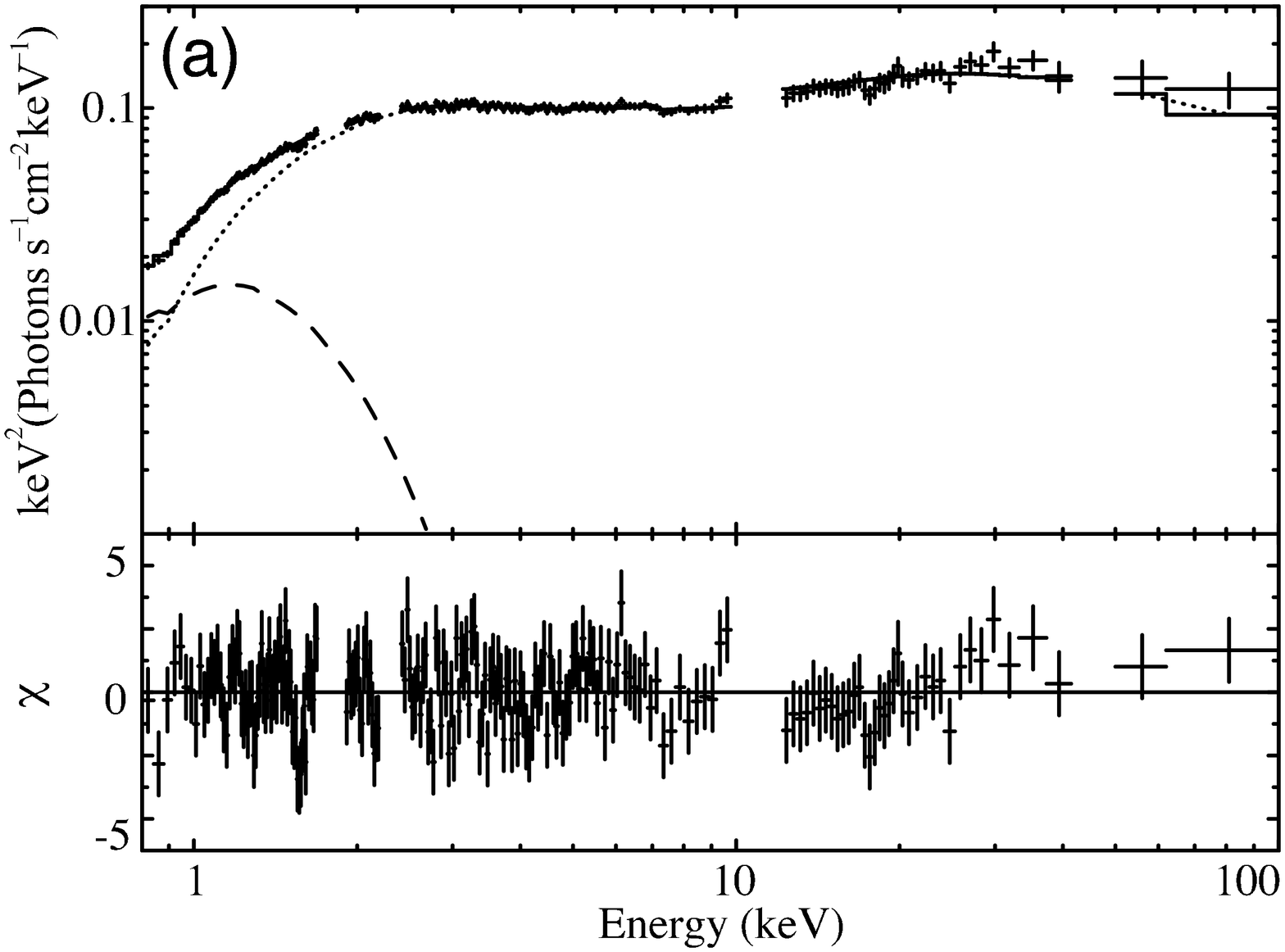}
		\FigureFile(80mm,80mm){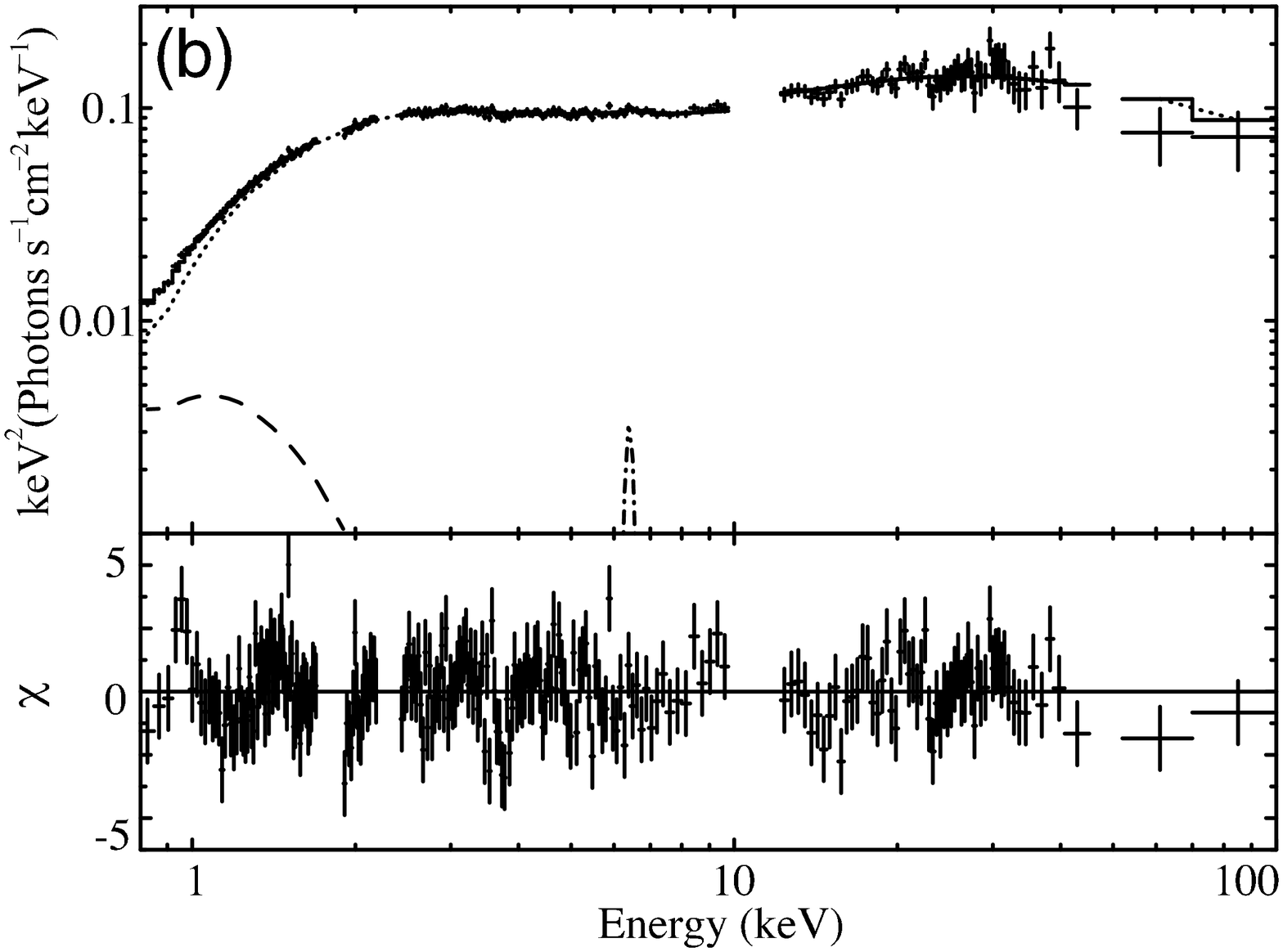}
	\end{center}
  \caption{Simultaneous fitting of the XIS, HXD-PIN, and HXD-GSO spectra in Obs$.$ 2 (panel a) and 
  Obs$.$ 4 (panel b) with {\tt diskBB+compPS(bbody)+gauss}.  Dashed, dotted, and dash-dotted lines show {\tt diskBB}, {\tt compPS}, and {\tt gauss} components, respectively.
  }
  			\label{fig_2ndand4th}
\end{figure*}

We hence fitted the Obs$.$ 2 and Obs$.$ 4 spectra with the same 
{\tt diskBB}+{\tt compPS}({\tt bbody})+{\tt gauss} model,
where the {\tt compPS} model \citep{Poutanen1996} incorporated disk reflection signals.
As already described in Paper 1, 
this is because the HXD spectra of these observations
bear a hump at 20--40 keV,
and a hint of Fe-K edge at $\sim 7.1$ keV,
both attributable to the reflection process.
This is reasonable also from a theoretical viewpoint,
because a certain fraction of the generated hard X-ray photons 
would inevitably hit the optically-thick outer part of the disk.
As shown in figure \ref{fig_2ndand4th} and table \ref{table_2ndand4th}, the two spectra
were both successfully reproduced by the model.
The obtained parameters, such as $R_\textmd{\scriptsize{in}}\sim$20 km and the blackbody radius (assuming spherical symmetry) 
$R_\textmd{\scriptsize{bb}}\sim$10 km, 
are the same within errors with those of Obs$.$ 3.
Using this model, the 0.8--100 keV luminosity was calculated as $2.5\times 10^{36}$ erg s$^{-1}$ for Obs$.$ 2 and $2.4\times 10^{36}$ erg s$^{-1}$ for Obs$.$ 4.
These are close to that of Obs$.$ 3, $2.9\times 10^{36}$ erg s$^{-1}$ (Paper 1).

In these analyses, we fixed the absorbing column density 
to $N_{\rm \scriptsize{H}} = 0.36 \times 10^{22}$ cm$^{-2}$ 
which was obtained in Paper 1 using the soft-state data.
This is based on the following two reasons.
One is that the soft-state (Obs$.$ 1) spectrum is brighter by 
an order of magnitude than those in the hard state (Obs$.$ 2--4),
yielding a higher accuracy of the absorption determination.
The other is that the soft X-ray ($\sim 2$ keV) 
spectral modeling in the soft state is considered to be 
much less ambiguous, than in the hard state
where the soft X-ray spectrum is affected by 
shapes of the seed photons for Comptonization,
and by the possible presence of an additional softer emission
(like the disk emission as seen above).

For reference, we examined 
whether the spectra of Obs$.$ 2---4
prefer a different value of  $N_{\rm \scriptsize{H}}$.
To avoid strong couplings between $N_{\rm \scriptsize{H}}$ 
and the disk emission parameters, $N_{\rm \scriptsize{H}}$ was allowed to vary freely,
but was constrained to be common among  Obs$.$ 2, Obs$.$ 3, and Obs$.$ 4.
This yielded  $N_{\rm \scriptsize{H}} = (0.38 \pm 0.05) \times 10^{22}$ cm$^{-2}$,
in good agreement with the  pre-assumed value  from Obs$.$ 1.
Furthermore, the fit results, including the presence of the additional soft emission,
have remained intact even when $N_{\rm \scriptsize{H}}$ was thus left free.
These results justify the use of  $N_{\rm \scriptsize{H}} = 0.36 \times 10^{22}$ cm$^{-2}$.


\begin{table*}[htb]
\caption{Parameters of the {\tt diskBB} plus {\tt compPS}({\tt bbody}) fit for Obs$.$ 2 and Obs$.$ 4.\footnotemark[$*$] }
\centering
\begin{minipage}{12cm}
\begin{tabular}{ccccc}\hline
Component & Paramater & Obs$.$ 2	& Obs$.$ 3 \footnotemark[$\dagger$]  & Obs$.$ 4 \\\hline
{\tt wabs} 	 & $N_\textmd{\scriptsize{H}}$ ($10^{22}$cm$^{-2}$) & \multicolumn{3}{c}{0.36 (fixed)} \\\hline
{\tt diskBB} & $T_\textmd{\scriptsize{in}}$ (keV) 					& $0.28_{-0.02}^{+0.03}$  & $0.28\pm0.02$ &  $0.23_{-0.04}^{+0.05}$	\\
	    	 & $R_\textmd{\scriptsize{in}}$ (km)\footnotemark[$\ddagger$]\footnotemark[$\S$]  &$19\pm 4$ & $21\pm 4$ & $18_{-6}^{+16}$
	    	 \\\hline {\tt compPS(bbody)}& $T_\textmd{\scriptsize{bb}}$ (keV) 			 & $0.52\pm 0.02$			& $0.51\pm 0.02$ & $0.49\pm 0.02$ 
	    	 \\
	     	 & $T_\textmd{\scriptsize{e}}$ (keV)  			 & $48_{-6}^{+7}$ 			& $35_{-5}^{+4}$ & $48\pm 6$ \\
	 		 & optical depth 					 			 & $2.0\pm 0.2$	 			& $\geq 2.5$ & $2.0\pm 0.2$  \\
	 		 & reflection ($2\pi$)					 		 & $0.6\pm 0.2$	 			& $0.6_{-0.1}^{+0.2}$ & $0.6\pm0.2$  \\
	      	 & $R_\textmd{\scriptsize{bb}}$  (km)\footnotemark[$\ddagger$]\footnotemark[$\S$] & $9\pm 2$		& $10\pm 2$  & $10\pm2$\\\hline
{\tt Gaussian}\footnotemark[$\|$] & EW (eV)\footnotemark[$\#$] 			 &	$< 16$		  & $22_{-16}^{+13}$		   & 		$9_{-8}^{+12}$	 \\\hline
Fit goodness & $\chi_\nu^2(\nu)$ & 1.22 (202) & 1.02 (183) & 1.21 (238)  \\\hline
Luminosity\footnotemark[$**$] & (erg s$^{-1}$)  & $2.5\times 10^{36}$ & $2.9\times 10^{36}$ erg s$^{-1}$ & $2.4\times 10^{36}$  \\\hline
\end{tabular}
\label{table_2ndand4th}
\footnotetext[$*$]{Errors represent 90\% confidence limits.}
\footnotetext[$\dagger$]{Referential values quoted from Paper 1.}
\footnotetext[$\ddagger$]{Calculated in the same way with Paper 1 (assuming the same distance of 5.2 kpc).}
\footnotetext[$\S$]{Also propagating the distance error of $\pm 0.7$ kpc \citep{Jonker2004}.}
\footnotetext[$\|$]{The centroid energy and width are fixed to 6.4 keV and 0.1 keV, respectively.}
\footnotetext[$\#$]{Equivalent width.}
\footnotetext[$**$]{In the 0.8--100 keV energy range.}
\end{minipage}
\end{table*}

\subsection{Obs$.$ 5}\label{ss_obs5}

\begin{figure*}[htb]
	\begin{center}
		\FigureFile(80mm,80mm){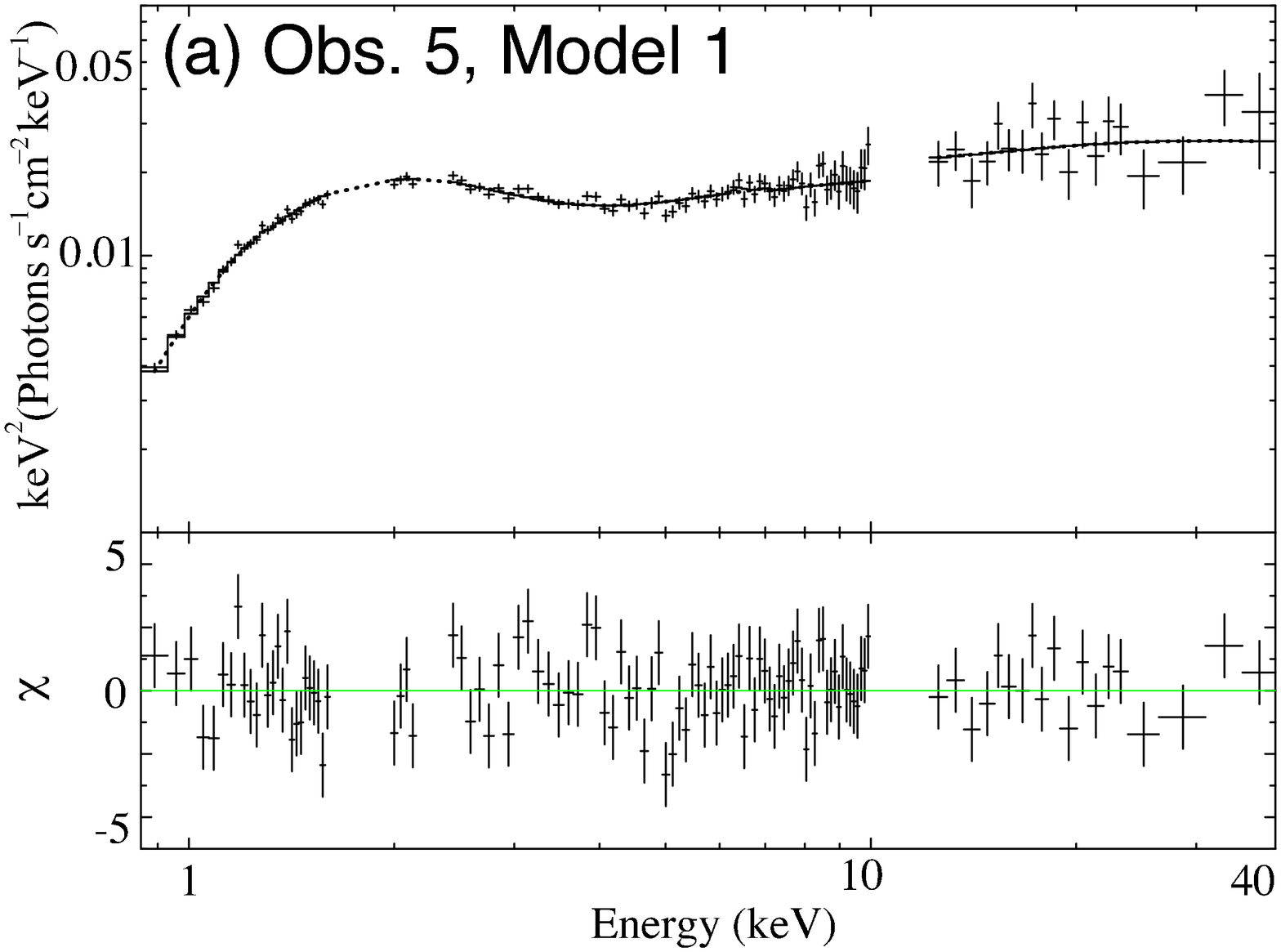}
		\FigureFile(80mm,80mm){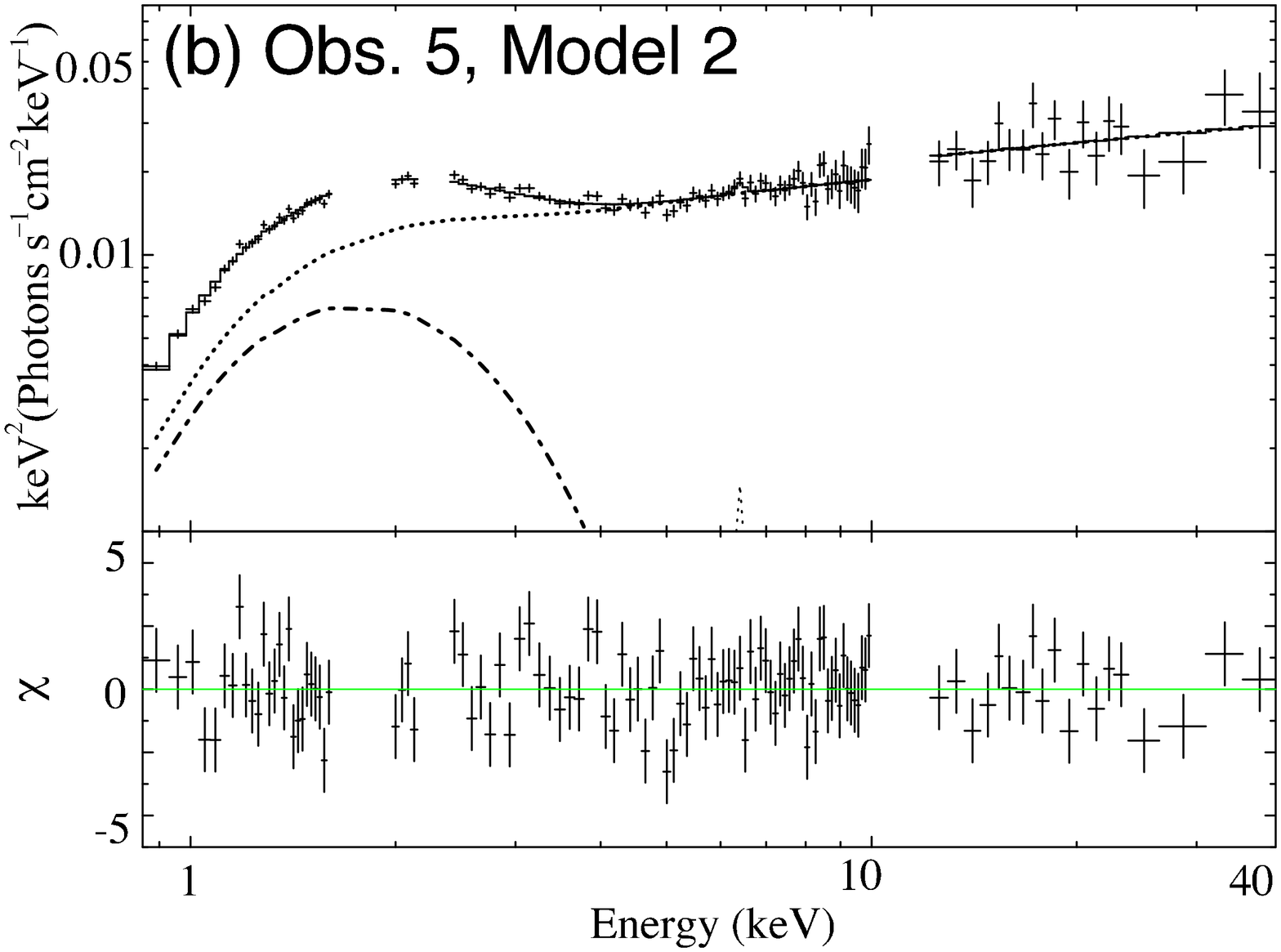}
	\end{center}
  \caption{Simultaneous fitting of the XIS and HXD-PIN spectra in Obs$.$ 5. (a) A fit with Model 1.
  			(b) A fit with Model 2, where the directly-seen {\tt bbody} is shown in dot-dashed curve while the Comptonized component in dotted curve.
			The composition of Model 1 and Model 2 is described in section \ref{ss_obs5}.
			}
  			\label{fig_5th}
\end{figure*}


\begin{table*}[htb]
\caption{Results of Model 1 and Model 2 fits for Obs$.$ 5.\footnotemark[$*$] }
\centering
\begin{minipage}{10cm}
\begin{tabular}{cccc}\hline
Component & Paramater & Model 1 & Model 2 \\\hline
{\tt wabs}	 & $N_\textmd{\scriptsize{H}}$ ($10^{22}$cm$^{-2}$) & \multicolumn{2}{c}{0.36 (fixed)} \\\hline
{\tt bbody}	 & $T_\textmd{\scriptsize{bb}}$ (keV) 		 					& - 	 	 & $0.39\pm 0.01$ \\
	   		 & $R_\textmd{\scriptsize{bb}}$ (km)\footnotemark[$\dagger$]\footnotemark[$\ddagger$]  & - 		 	 & $5\pm 1$ \\\hline
{\tt compPS(bbody)}& $T_\textmd{\scriptsize{bb}}$ (keV)\footnotemark[$\S$] & $0.40\pm 0.01$		 & $0.39\pm 0.01$ \\
	      			& $T_\textmd{\scriptsize{e}}$ (keV)  & $120_{-10}^{+20}$ 		 	 & $48_{-4}^{+65}$ \\
	 			    & optical depth 					 & $0.81_{-0.04}^{+0.05}$	 	 & $>3$ \\
	      			& $R_\textmd{\scriptsize{bb}}$  (km)\footnotemark[$\ddagger$] & $7\pm 1$		 & $6\pm 1$ \\\hline
Fit goodness & $\chi_\nu^2(\nu)$ & 1.22 (110) & 1.20 (109) \\\hline
Luminosity\footnotemark[$\|$] & (erg s$^{-1}$) &  $4.8\times 10^{35}$  &  $5.2\times 10^{35}$ \\\hline
\end{tabular}
\label{table_5th}
\footnotetext[$*$]{Errors represent 90\% confidence limits.}
\footnotetext[$\dagger$]{After applying the corrections described in the text, and assuming a distance of 5.2 kpc.}
\footnotetext[$\ddagger$]{Also propagating the distance error of $\pm 0.7$ kpc \citep{Jonker2004}.}
\footnotetext[$\S$]{Tied to that of directly-seen {\tt bbody} when Model 2.}
\footnotetext[$\|$]{In the 0.8--100 keV energy range.}
\end{minipage}
\end{table*}

The spectrum of Obs$.$ 5 looks similar in shape to those of Obs$.$ 2 through Obs$.$ 4,
except that it is clearly concave around 4 keV in the $\nu F\nu$ form.
On this occasion, the 0.8--40 keV intensity was $\sim 18\%$ of those in the preceding three observations (see figure \ref{table_2ndand4th}).
Therefore, we assume that the spectral composition employed in section \ref{ss_obs2and4} is still valid,
possibly with some changes in its parameters. 
In this context, there are two possible model modifications that make the spectrum concave, 
in other words, make the soft thermal emission stand out.
One is that the coronal optical depth decreased due to the reduced accretion rate,
and the number of non-scattered photons increased.
The other is that some fraction of the seed blackbody became directly visible, not covered by the corona.
Hence we tried two models representing these conditions, namely {\tt compPS}({\tt bbody}) and {\tt bbody}+{\tt
compPS}({\tt bbody}), hereafter Model 1 and Model 2, respectively, 
where the model name in the parenthesis indicates the seed photon source.
In Model 2, the two blackbodies, one directly visible where the other supplying seed photons,
are constrained to have the same temperature.

The above two models are both natural extensions/modifications
to that employed in section \ref{ss_obs2and4} to describe Obs$.$ 2--4.
Particularly, Model 1 is essentially the same as that given in table 2,
except that the optically-disk emission was omitted.
This exclusion is justified,
because the factor $\sim 5$ intensity decrease from Obs$.$ 2--4 to Obs$.$ 5 
would reduce the inner disk temperature by $\sim 40\%$ 
if $R_\textmd{\scriptsize{in}}$ remained the same
(and much more if the disk further retreated back),
thus causing the disk emission to fall 
below the lower energy boundary of Suzaku.
Model 2 is motivated by a consideration
that the clear spectral hump below $\sim 4$ keV could 
be  attributed to another softer optically-thick emission,
but the disk would be too cool to explain this hump.
Because the spectrum does not bear iron-edge feature, the reflection component was not incorporated in either Model 1 or Model 2.
The results to be reported below remains the same within errors,
even if we include the reflection component with its solid angle
fixed at $1.2 \pi$ (table \ref{table_2ndand4th}).

As shown in figure \ref{fig_5th} and table \ref{table_5th},
the spectrum was well fit by either model,
without the disk reflection signals (included in {\tt compPS}).
In Model 1, some seed photons from the neutron-star surface are scattered by energetic electrons 
in the corona with a temperature of $T_\textmd{\scriptsize{e}} \sim 120$ keV, 
while the relatively low optical depth, $\tau\sim 0.8$, allows the remaining fraction of seed photons 
to escape without any scattering.
This results in the zero-scattered blackbody component at $< 2$ keV and the $\Gamma\sim 2$ power-law-like
Comptonization component with a $y$-parameter of $\sim 0.8$.
In Model 2, the spectrum is reproduced by a directly-seen blackbody which accounts for the hump in $\lesssim 3$ keV,
and a hard Comptonization component.
The best-fit Model 2 implies nearly the same incident spectrum as Model 1, yielding a comparable fit goodness.
However, the Model 2 result differs from that with Model 1, in that the optical depth, $\tau > 3$,
is much higher (not to produce significant leak-through photons), and the best-fit coronal temperature, 
$T_\textmd{\scriptsize{e}} \sim 50$ keV, is much lower (to compensate for the higher $\tau$).
The two solutions would have been distinguished if the source were detected 
in energies of $>40$ keV with the HXD-GSO.

In the Model 2 solution, the directly visible blackbody has a radius of $5\pm 1$ km,
while that which supplies Compton seed photons has a comparable radius of $6\pm 1$ km.
This means that about half the blackbody-emitting area is covered by the corona.
When the two radii are summed up in quadrature, we obtain an overall blackbody radius of $R_\textmd{\scriptsize{bb}}=7\pm 1$ km 
which agrees well with that from Model 1.
Thus, regardless of the model ambiguity, the data yields $R_\textmd{\scriptsize{bb}} = 7\pm 1$ km,
which is significantly smaller than those measured in Obs$.$ 2, Obs$.$ 3, and Obs$.$ 4 ($\sim 10$ km).


Although the source was detected only up to 40 keV, 
it is convenient to keep using the luminosity in the 0.8--100 keV range to be compared with the earlier observation.
We therefore extrapolated the two best-fit models, and obtained the 0.8--100 keV luminosity as 
table \ref{table_5th}.
Thus, the high-energy extrapolations of the two models agree, within $\sim 10$\%, on the wide-band luminosity.

\subsection{Obs$.$ 6 and 7}
\begin{figure*}[htb]
	\begin{center}
		\FigureFile(80mm,80mm){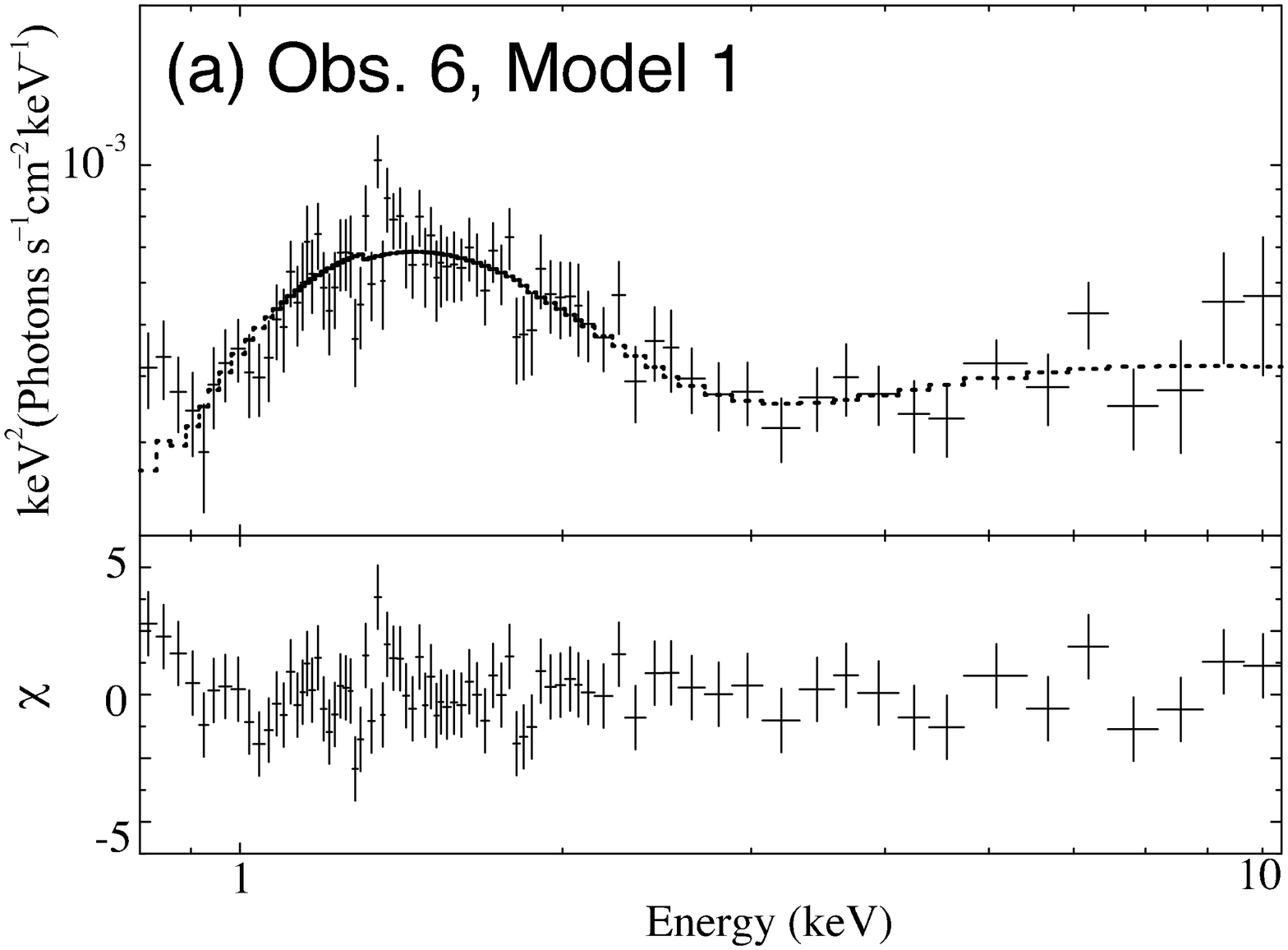}
		\FigureFile(80mm,80mm){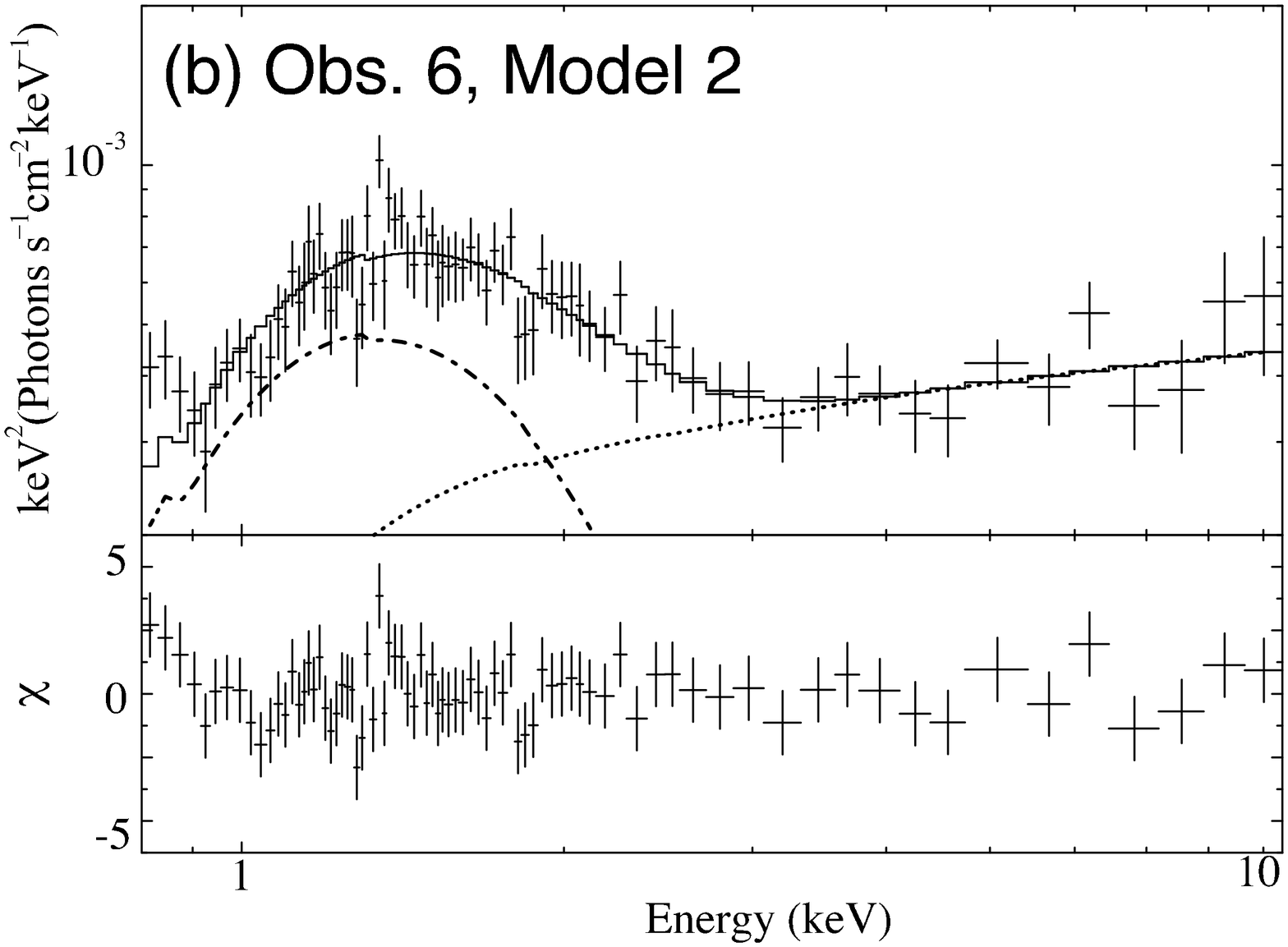}
		\FigureFile(80mm,80mm){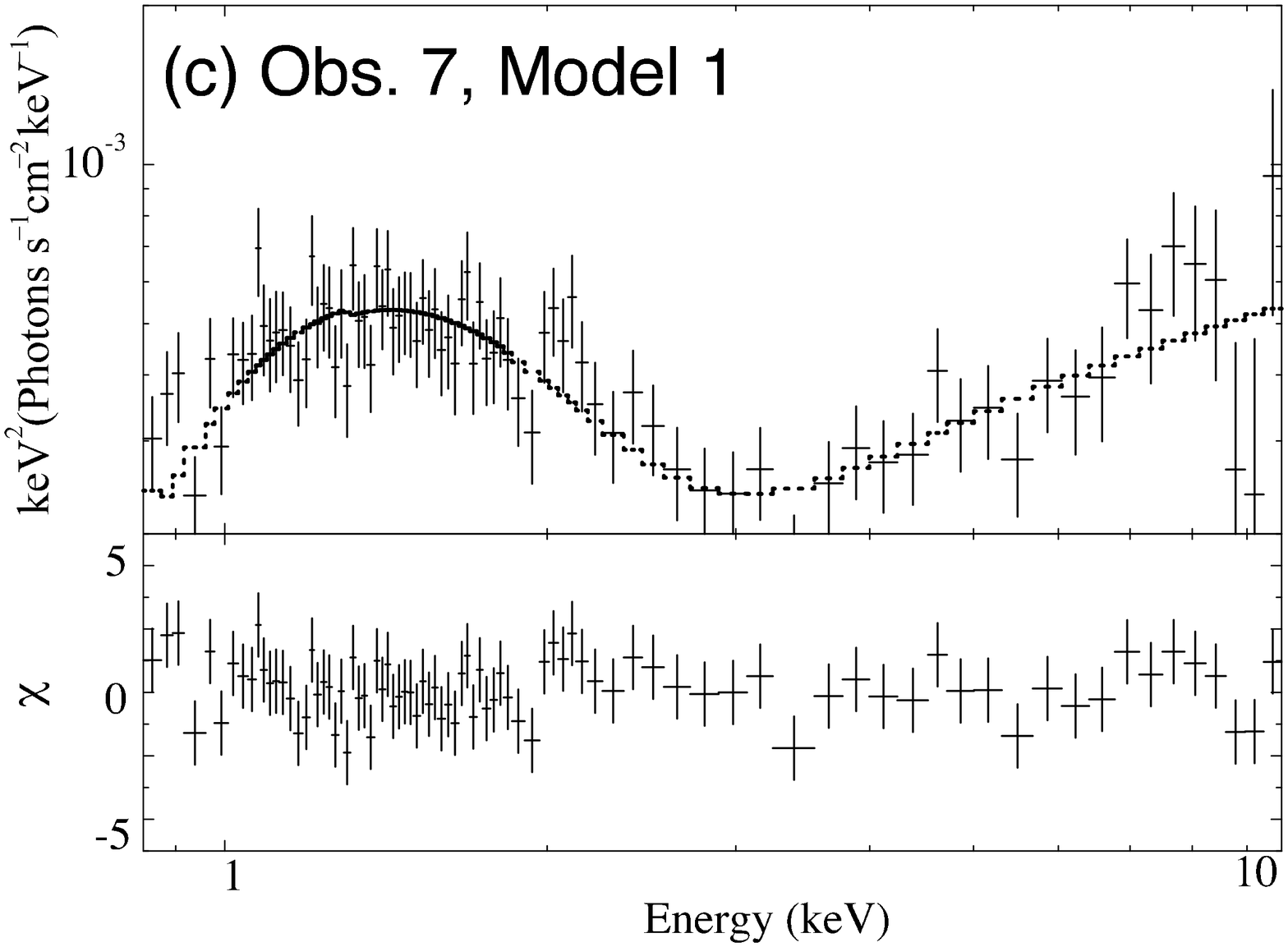}
		\FigureFile(80mm,80mm){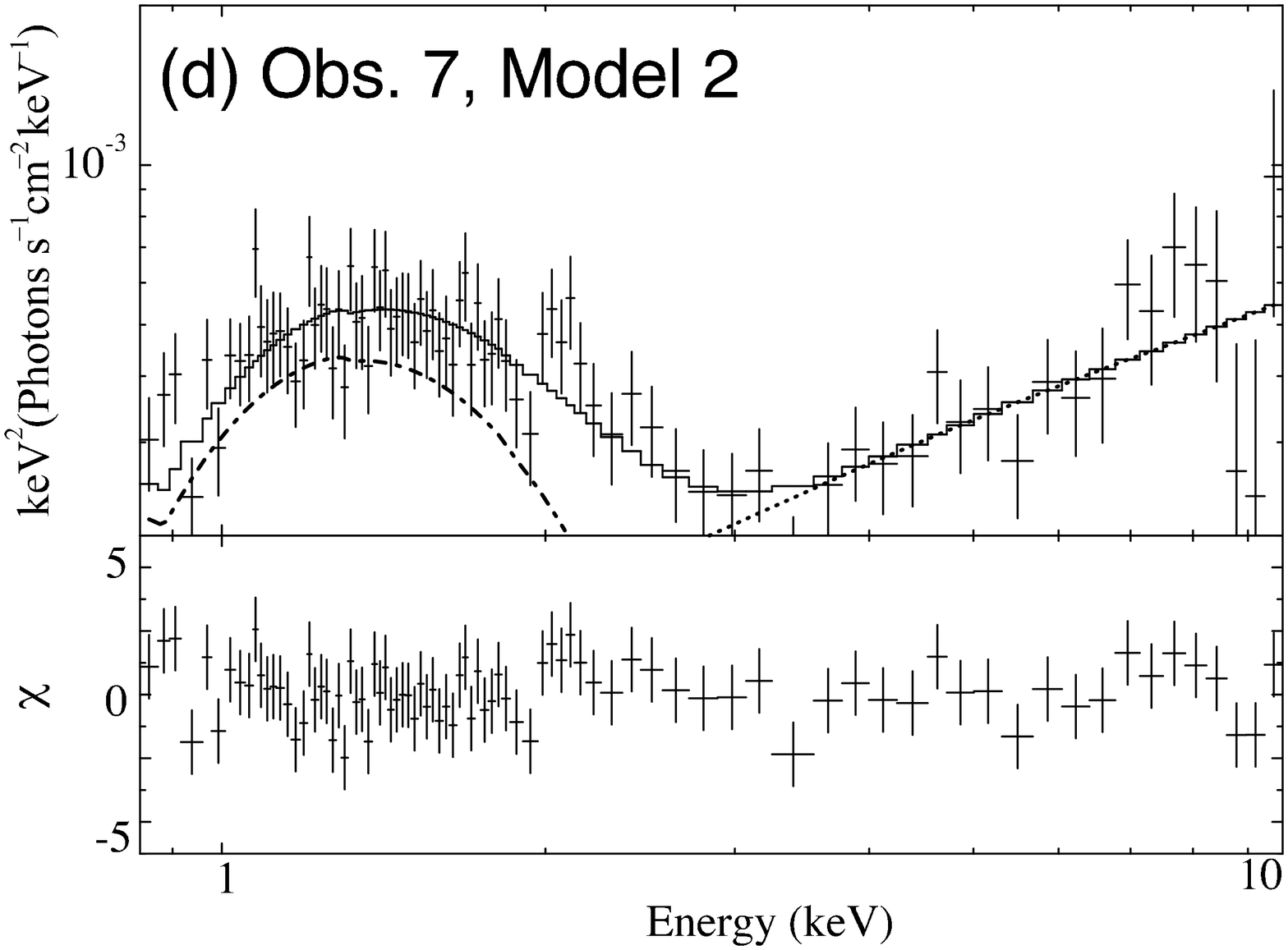}
	\end{center}
  \caption{Model fittings of the XIS spectra in Obs$.$ 6 and 7.
  				Top two panels (a and b) are for Obs$.$ 6, while bottom two (c and d) are for Obs$.$ 7.
				Left two panels (a and c) use Model 1, while right two (b and d) Model 2 where the dot-dashed curve represents {\tt bbody} and the dotted curve {\tt compPS}.
  	}
  			\label{fig_6thand7th}
\end{figure*}

\begin{table*}[htb]
\caption{Parameters of Model 1/2 fit for Obs$.$ 6 and 7.\footnotemark[$*$] }
\centering
\begin{minipage}{12cm}
\begin{tabular}{cccccc}\hline
Component   & Paramater   & \multicolumn{2}{c}{Obs$.$ 6}		& \multicolumn{2}{c}{Obs$.$ 7} \\
		  &			 	  & Model 1					  & Model 2  & Model 1 & Model 2 \\\hline
{\tt wabs} & $N_\textmd{\scriptsize{H}}$ ($10^{22}$cm$^{-2}$) & \multicolumn{2}{c}{0.36 (fixed)} & \multicolumn{2}{c}{0.36
(fixed)}\\\hline 
{\tt bbody} & $T_\textmd{\scriptsize{bb}}$ (keV) 		 						 & -	 	 & $0.27\pm 0.02$ & - & $0.27\pm 0.02$ \\
	    & $R_\textmd{\scriptsize{bb}}$ (km)\footnotemark[$\dagger$]\footnotemark[$\ddagger$] & -  & $3\pm 1$ 	& - & $3\pm 1$
	    \\\hline {\tt compPS(bbody)}& $T_\textmd{\scriptsize{bb}}$ (keV)\footnotemark[$\S$] 		 & $0.27\pm 0.01$	& $0.27\pm 0.02$ &
$0.27\pm 0.02$&$0.27\pm 0.02$\\
	      			& $T_\textmd{\scriptsize{e}}$ (keV)  						 & $160_{-40}^{+50}$ 	& $50_{-10}^{+80}$ & $300_{-100}^{+200}$	&
	      			$150_{-70}^{+230}$ \\
	 			    & optical depth 					 						 & $0.30_{-0.05}^{+0.06}$	& $>3$ 	& $0.24_{-0.03}^{+0.04}$& $>3$ \\
	      			& $R_\textmd{\scriptsize{bb}}$  (km)\footnotemark[$\ddagger$]& $3\pm 1$ 			& $2_{-1}^{+2}$	& $3\pm 1$ & $2_{-1}^{+3}$
	      			\\\hline Fit goodness & $\chi_\nu^2(\nu)$ 	& 0.91 (76) & 0.91 (75)	& 0.86 (81) & 0.87 (80) \\\hline
Luminosity\footnotemark[$\|$] & (erg s$^{-1}$) &  $1.2\times 10^{34}$  &  $1.5\times 10^{34}$ &  $1.5\times 10^{34}$  &  $2.8\times 10^{34}$ \\\hline
\end{tabular}
\label{table_6thand7th}
\footnotetext[$*$]{Errors represent 90\% confidence limits.}
\footnotetext[$\dagger$]{After applying the corrections described in the text, and assuming a distance of 5.2 kpc.}
\footnotetext[$\ddagger$]{Also propagating the distance error of $\pm 0.7$ kpc \citep{Jonker2004}.}
\footnotetext[$\S$]{Tied to that of directly-seen {\tt bbody} when the coronal covering fraction is left free.}
\footnotetext[$\|$]{In the 0.8--100 keV energy range.}
\end{minipage}
\end{table*}

Among the seven observations, Obs$.$ 6 and 7 covered the faintest phase,
wherein the source intensity had already decreased rapidly.
The 0.8--10 keV intensity (table \ref{table_obsproperty}) on these two occasions are
no higher than $\sim 5$\% of that in Obs$.$ 5,
and the spectra on these occasions are much more clearly concave at $\sim 3$ keV than that of Obs$.$ 5.
Nevertheless, the spectra are still quantitatively similar in shape to that of Obs$.$ 5.
Therefore, we proceeded to fit these spectra with a blackbody plus its Comptonization, namely, the same 
two alternative models as used in Obs$.$ 5 which in turn are both natural extensions to that for Obs$.$ 2--4.

As shown in figure \ref{fig_6thand7th} (panels a and b) and table \ref{table_6thand7th}, 
the Obs$.$ 6 spectrum was successfully reproduced by either model like in section \ref{ss_obs5}.
According to Model 1, the coronal temperature remained approximately the same as in Obs$.$ 5,
but the coronal optical depth decreased from $\sim 0.8$ to $\sim 0.3$.
This low optical depth enables more blackbody photons to pass through the corona without any scattering, 
to make the zero-scattered component stand out at $\sim 0.27$ keV and make the spectrum look more concave.
In Model 2, the coronal parameters turned out to be consistent within errors with those of Obs$.$ 5.
What makes the spectrum more concave is the change of coronal covering fraction, namely, 
ratio between the seed photon area to the total (seed plus direct) blackbody area.
In fact, from Obs$.$ 5 to Obs$.$ 6,
the ratio changed from $(6 {\rm km})^2/ \{ (6 {\rm km})^2 + (5 {\rm km})^2 \}\sim 0.6$ to $(2 {\rm km})^2/ \{ (2 {\rm km})^2 + (3 {\rm km})^2 \}\sim 0.3$ (though the errors are large), so that the direct blackbody 
component became stronger.

As shown in figure \ref{fig_6thand7th} (panels c and d), the Obs$.$ 7 spectrum was also well reproduced by either Model 1 or Model 2.
The more prominent concaveness of the Obs$.$ 7 spectrum, than that of Obs$.$ 6, 
can be explained as a further change of the model parameters, 
from Obs$.$ 5 through Obs$.$ 6, and to Obs$.$ 7.
If, e.g., Model 1 is employed, the optical depth further decreased to $\sim 0.24$  (table \ref{table_6thand7th}).
If, instead, Model 2 is employed, the spectral change from Obs$.$ 6 to Obs$.$ 7 is mainly attributed to the increase (though with large errors)
in the coronal electron temperature, which made the Comptonization continuum harder.

By extrapolating the best-fit two models, like in section \ref{ss_obs5}, we calculated the luminosity in the 0.8--100 keV band.
The results are shown in table \ref{table_6thand7th}.



\section{discussion}
\subsection{Spectral modeling}
In the present study, we analyzed five wide-band Suzaku spectra of Aql X-1 (the 2nd, 4th, 5th, 6th, 
and 7th) among the seven observations conducted over decay phase of the 2007 September-October outburst.
If the individual 0.8--100 keV luminosities (section \ref{s:spectral_analysis}; either directly measured or extrapolated)
are normalized to the Eddington luminosity of $1.8\times 10^{38}$ erg s$^{-1}$ for an object of $1.4M_\odot$,
we obtain the Eddington ratios of 0.077 for Obs$.$ 1, $\sim 0.013$ for Obs$.$ 2, 3, and 4, $\sim 2.7\times 10^{-3}$ for Obs$.$ 5,
and $\sim (0.74-1.1)\times 10^{-4}$ for the last two observations.
Thus, during these observations, the luminosity changed by $\sim 3$ orders of magnitude,
and the spectra (figure \ref{fig_allobs}) changed drastically meanwhile.

Like in Obs$.$ 3 already reported in Paper 1, Obs$.$ 2 and Obs$.$ 4  caught the source in the typical hard state,
wherein the spectra were reproduced by a strongly Comptonized blackbody plus a weaker disk blackbody.
The innermost disk radii, $19\pm 4$ km (Obs$.$ 2) and $18_{-6}^{+16}$ km (Obs$.$ 4), and the blackbody radii, 
$9\pm 2$ km (Obs$.$ 2) and $10\pm 2$ km (Obs$.$ 4), are all consistent with those of Obs$.$ 3.
Thus, at these luminosities, the standard disk is truncated at a radius of $\sim 20$ km,
and the accreting matter then transforms into an optically-thin, geometrically-thick coronal flow.
As discussed in Paper 1, this flow is considered to accrete almost spherically onto the neutron star and
make its entire surface emit the blackbody photons with a temperature of $T_\textmd{\scriptsize{bb}}\sim 0.5$ keV and
a radius of $R_\textmd{\scriptsize{bb}}\sim 10$ km (table \ref{table_2ndand4th}).
These photons are then Compton scattered by the succeeding coronal flows.

The spectra of  Obs$.$ 5$-$7 are reproduced by a blackbody and its Comptonization,
without the disk component which is probably below the XIS energy band.
In these three faintest data sets, it is ambiguous, however, whether the blackbody source is covered entirely by 
a relatively thin ($\tau = 0.2\sim 0.8$) and very hot ($T_\textmd{\scriptsize{e}} \gtrsim 100$ keV) corona (Model 1),
or partially by a thicker ($\tau > 3$) and cooler ($T_\textmd{\scriptsize{e}} \sim 50$ keV) one (Model 2).
The characteristic hump below $\sim 3$ keV seen in these spectra
can be explained either by those blackbody photons which escaped through the thin corona
without scattering (Model 1), or emission from those regions on the neutron star which are not covered by the thick corona (Model 2).

Given the model ambiguity in the dimmest three data sets,
we may ask ourselves whether the Obs$.$ 2$-$4 data,
which have so far been analyzed with Model 1 (plus {\tt diskBB}),
can also be described with Model 2 (plus {\tt diskBB}).
To see this, we fitted the Obs$.$ 2$-$4 spectra 
with a model {\tt diskBB} + {\tt compPS} +{\tt bbody} +{\tt gauss},
wherein the additional blackbody component, 
representing surface emission without coronal obscuration,
was constrained to have the same temperature as the Compton seed photons.
Then, the directly visible blackbody was found to little contribute,
with the radius of $<3$ km, $<2$ km,  and $ <4$ km
from Obs$.$ 2, Obs$.$ 3 and Obs$.$ 4, respectively.
In none of these cases, the fit improvement was significant.
Therefore, these data sets prefer Model 1, in agreement with Paper 1 in which the corona was
concluded to be approximately spherical.

\subsection{Changes of the blackbody radius}
\begin{figure}[htbp]
	\begin{center}
		\FigureFile(88mm,120mm){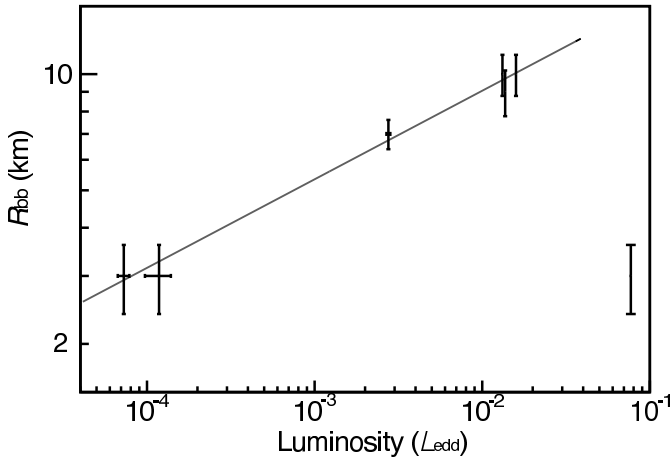}
	\end{center}
\caption{Luminosity dependence of the blackbody radius $R_{\rm bb}$ (tables \ref{table_2ndand4th} to \ref{table_6thand7th}) of Aql X-1 from the seven observations.
The $R_{\rm bb}$ values are calculated as for Model 1, and are the same as those in Model 2.
The luminosity is calculated or estimated in the 0.8--100 keV range, 
and normalized by the Eddington luminosity for the neutron-star mass of $1.4M_\odot$, 
while the errors include Model 1/Model 2 systematic differences. 
 The solid line shows the best power-law fit excluding the Obs$.$ 1 data point.
}
\label{fig:Rbbfig}
\end{figure}

An important source property revealed by analyzing the seven data sets is luminosity-dependent changes in $R_\textmd{\scriptsize{bb}}$,
which hold regardless of the Model 1/Model 2 ambiguity.
This is evident in figure \ref{fig:Rbbfig}, where $R_\textmd{\scriptsize{bb}}$ is plotted against the Eddington-normalized
0.8--100 keV luminosity.
As discussed in Paper 1, the small value of $R_\textmd{\scriptsize{bb}}$ in Obs$.$ 1 is considered to mean that the accretion 
in the soft state occurs only onto an equatorial region of the neutron star.
As $L$ decreased to $\sim 10^{36}$ erg s$^{-1}$ (Obs$.$ 2$-$4) and the source made a transition into the hard state, 
the emission radius increased to  $R_\textmd{\scriptsize{bb}}\sim 10$ km, which is consistent with the neutron-star radius 
$R_\textmd{\scriptsize{NS}}$.
This is thought to be a consequence of the nearly spherical coronal flow (Paper 1).
As the source became still dimmer,  $R_\textmd{\scriptsize{bb}}$ started decreasing again, down to $\sim 3$ km at a luminosity 
of $\sim 10^{34}$ erg s$^{-1}$.
As shown in figure \ref{fig:Rbbfig} by a solid line, the dependence of $R_\textmd{\scriptsize{bb}}$ on $L$ (in Obs$.$ 2 to Obs$.$ 7) is approximated as
$R_\textmd{\scriptsize{bb}} \propto L^{0.23\pm 0.03}$.

The decrease of $R_\textmd{\scriptsize{bb}}$ towards lower luminosity is presumably 
due to some collimation of the optically-thin accretion flow. This, in turn, is likely to be caused by the
emergence of a neutron-star magnetosphere which is usually negligible when the source is luminous.
To evaluate this idea, let us recall that the Alfven radius $R_{\rm A}$ (for spherical accretion) depends on the mass accretion rate $\dot{M}$ as
\begin{eqnarray}
R_{\rm A} &=& 3.0\times 10^3 \left( \frac{\dot{M}}{10^{14} \; {\rm  kg/s}} \right)^{-2/7} \left( \frac{B}{10^{12} \; {\rm  G}} \right)^{4/7} ({\rm km})\nonumber\\
 &=& 1.9\times 10^3 \left( \frac{ L }{10^{38} \; {\rm  erg/s}} \right)^{-2/7} \left( \frac{B}{10^{12} \; {\rm  G}} \right)^{4/7} ({\rm km})
\label{eq:alfvenradius}
\end{eqnarray}
(Elsner \& Lamb 1977),
where $B$ is the neutron-star magnetic field; here we assumed the neutron-star mass and radius of $M_{\rm NS}=1.4M_\odot$ and $10$ km, respectively, and the magnetic field of dipolar configuration.
We may consider that $R_\textmd{\scriptsize{bb}}$ starts decreasing in figure \ref{fig:Rbbfig} at an Eddington ratio of $\sim 10^{-2}$, 
or $L\sim 1\times 10^{36}$ erg s$^{-1}$,
because the magnetic field starts to affect the accretion flow,
or equivalently, $R_\textmd{\scriptsize{A}}$ becomes comparable to or larger than $R_\textmd{\scriptsize{NS}}$.
This, together with equation \ref{eq:alfvenradius}, yields $B\sim 1\times 10^7$ G.

Since the rotation frequency of Aql X-1 is considered as $\nu=$550 Hz (Zhang et al$.$ 1998; Casella et al$.$ 2008), 
the corotation radius is calculated as
\begin{equation}
R_{\rm co} = \left(\frac{GM_{\rm NS}}{4\pi^2 \nu^2}\right)^{1/3} = 25\; {\rm km}.
\end{equation}
Employing the field strength obtained above, $R_\textmd{\scriptsize{A}}$ is then inferred to exceed $R_{\rm co}$ at $L \lesssim 10^{35}$ erg s$^{-1}$.
At such low luminosities (like in Obs$.$ 6 and Obs$.$ 7), the accreting matter will be 
strongly affected by the magnetic pressure, and will start extracting
the angular momentum from the neutron star before it reaches $r\sim R_\textmd{\scriptsize{co}}$.
This effect, often called ``propeller effect" \citep{Lamb1973, Matsuoka2012}, can explain the large luminosity drop from Obs$.$ 5 to Obs$.$ 6 and Obs$.$ 7.


\subsection{Information from the coronal optical depth}
Although $R_{\rm bb}$ was found to 
reduce as $\dot{M}$ decreases,
the effect could be some artifact; 
e.g. due to radiative transfer in the NS atmosphere (e.g. Heinke et al$.$ 2006).
As an independent examination, 
let us consider here the luminosity dependence of the coronal optical depth $\tau$. 
After equation (6) of Paper I, 
the mass continuity condition along the radius $r$ can be written as
\begin{equation}
 \dot{M} = S(r) \; g(r) \;v(r)\; \mu m_{\rm p} \; n(r) ~,
\label{eq:mass_continuity}
\end{equation}
where $\mu \sim 1.2$ is the mean molecular weight of ions,
$m_{\rm p}$ is the proton mass,
$n$ is the flow density,
$S$ is the flow cross section ($=4 \pi r^2$ if spherical),
$v \propto r^{-1/2}$ is the free fall velocity,
and $g$ denotes possible deviation of 
the actual inflow velocity from $v$ (Paper 1).
Assuming that $\dot{M}$ is constant,
that $S(r )$ approximately scales as $\propto r^2$
(nearly radial flow),
and that $g(r)$ depends only weakly on $r$,
equation (\ref{eq:mass_continuity}) yields
\begin{equation}
 n(r)=   \left( \dot{M} / \mu m_{\rm p}S_0  g v_0 \right) r^{-1.5} ~,
\end{equation}
where subscript 0 specifies values on the neutron-star surface,
$r=R_{\rm NS}$.
Like equation (7) in Paper I, 
we may integrate this from $R_{\rm NS}$
to an outer coronal radius $R_{\rm c}$, to express the optical depth as
\begin{equation}
\tau \propto \left( \dot{M}/gS_{\rm 0} \right)  \left(R_{\rm NS}^{-0.5} - R_{\rm c}^{-0.5} \right)~~.
\end{equation}
Further assuming $R_{\rm c} \gg R_{\rm NS}$ 
and a negligible dependence of $g$ on $\dot{M}$,
we finally obtain 
\begin{equation}
S_{\rm 0} \propto \dot{M}/\tau \propto L/\tau
\label{eq:flow_cross_sec}
\end{equation}
where $\dot{M}$ was replaced with the 0.8--100 keV luminosity $L$.

\begin{figure}[htbp]
	\begin{center}
		\FigureFile(75mm,120mm){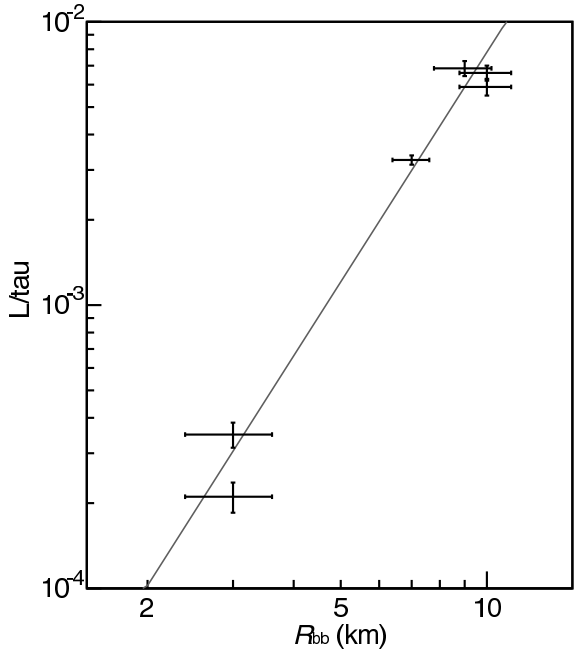}
	\end{center}
\caption{A relation between the blackbody radius $R_{\rm bb}$ and the quantity $L/\tau$ for Model 1.
The data point of Obs$.$ 1 is excluded here.
The solid line shows the best-fit power law of slope $2.7$.
}
\label{fig:Rbb-L_tau}
\end{figure}

To see the behavior of this $S_0$,
the values of $L/\tau$ for Model 1 were calculated from table 2, table 3, and table 4,
and are  shown in figure \ref{fig:Rbb-L_tau} against $R_{\rm bb}$.
The four data points indicate a very tight relation as
$L/\tau \propto R_{\rm bb}^{2.7 \pm 0.4}$.
Combined with equation (\ref{eq:flow_cross_sec}),
this implies $S_{\rm 0} \propto R_{\rm bb}^{2.7 \pm 0.4}$, which is close to the $S_{\rm 0} \propto R_{\rm bb}^{2}$ scaling
expected if $R_{\rm bb}$ is correctly estimated.
Further incorporating the scaling of figure \ref{fig:Rbbfig}, we find $S_{\rm 0} \propto L^{0.6\pm 0.1}$,
which reconfirms that the flow becomes indeed constrained onto a limited surface area as $L$ decreases.
In other words, $\tau$ decreased only by less than an order of magnitude
as the luminosity became lower by more than 2 orders of magnitude
(from Obs$.$ 2$-$4 to Obs$.$ 5$-$7),
because the flow became more strongly funneled onto 
limited regions (most likely the magnetic poles).

If, instead, applying the same method to Model 2 (in Obs$.$ 5$-$7) by substituting $3\pm 1$ for the saturated $\tau$,
the dependence becomes $L/\tau \propto R_{\rm bb}^{4.5 \pm 0.5}$, of which the power index is too much larger than the normal value 2 ($S_{\rm 0} \propto R_{\rm bb}^2$).
Through these arguments, we consider that Model 1 is more preferable than Model 2, 
in agreement with the inference made at the end of subsection 4.1.
The configuration of Model 1 is still possible even when the coronal flow is funneled onto the magnetic poles,
because the flow would be rather spherical at $r>R_{\rm A}$.
This means that Model 1 (sometimes with an addition of {\tt diskBB})
can describe the broad-band spectrum of Aql X-1 in the hard state from Obs$.$ 2 through Obs$.$ 7,
over $\sim 2$ orders of magnitude in luminosity.
However, the reality could be an intermediate condition between Model 1 and Model 2.


\subsection{Accretion geometry}
Summarizing the results obtained so far, 
the $\dot{M}$-dependent changes in the accretion geometry of an NS-LMXB may be described in the following manner, 
in terms of the four radii, $R_{\rm A}$, $R_{\rm co}$, $R_{\rm in}$, and $R_{\rm NS}$ (Matsuoka \& Asai 2012).
\begin{enumerate}
\item When $\dot{M}$ is high and hence $R_\textmd{\scriptsize{A}} \ll R_\textmd{\scriptsize{co}}$, the gravity is dominant in the accretion (corresponding to Obs$.$ 1).  The optically-thick disk continues down to $R_\textmd{\scriptsize{in}} \sim R_\textmd{\scriptsize{NS}}$.

\item As $\dot{M}$ decreases, the system makes a transition into the hard state. Since $R_\textmd{\scriptsize{A}} < R_\textmd{\scriptsize{co}}$ and $R_\textmd{\scriptsize{A}} \lesssim R_\textmd{\scriptsize{NS}}$,
the gravity is still dominant, but the disk becomes truncated at $R_\textmd{\scriptsize{in}} \gg R_\textmd{\scriptsize{NS}}$, 
and the matter accretes nearly spherically onto the neutron star (Obs$.$ 2$-$4). 

\item When $\dot{M}$ decreases to make $R_\textmd{\scriptsize{NS}} < R_\textmd{\scriptsize{A}} < R_\textmd{\scriptsize{co}}$,
the effect of magnetic field begins to stand out.  
The flow becomes anisotropic, and funneled onto the magnetic poles (Obs$.$ 5). 

\item As $\dot{M}$ further decreases, the magnetic field becomes more dominant, resulting in a condition of $R_\textmd{\scriptsize{co}} < R_\textmd{\scriptsize{A}}$.
The propeller effect (Lamb et al$.$ 1973) hampers the accretion, thus causing a rapid luminosity decrease.
However, a small fraction can still fall along the magnetic field lines onto a small area at the magnetic poles (Obs$.$ 6 and Obs$.$ 7).  
\end{enumerate}


\section{Conclusion}

Through an analysis of the seven Suzaku data sets of Aquila X-1 tracing an outburst decay, 
we obtained spectra at various luminosities (from $\sim 10^{37}$ erg s$^{-1}$ to $\sim 10^{34}$ erg s$^{-1}$), 
and carried out model fitting to each of them.
Consequently we obtained the results as below.
\begin{enumerate}
\item Like Obs$.$ 3 which we reported in Paper 1, 
Obs$.$ 2 and Obs$.$ 4 had luminosities of $L\sim 10^{36}$ erg s$^{-1}$, and showed typical characteristics of the hard state.
The spectra are composed of a Comptonized blackbody emission from the whole neutron-star surface and a softer emission from a truncated disk.

\item The spectra of Obs$.$ 5, Obs$.$ 6, and Obs$.$ 7 were reproduced either by a Comptonized blackbody (Model 1) or that with directly-seen blackbody (Model 2).
The obtained radius of the blackbody emission was significantly smaller than 10 km and decreased along with the luminosity,
regardless of the model ambiguity.

\item We clarified luminosity dependence of $R_\textmd{\scriptsize{bb}}$, 
which is approximated as $R_\textmd{\scriptsize{bb}} \propto L^{0.23\pm 0.03}$.
This can be considered as an evidence the emergence of a neutron-star magnetosphere,
which confines the accretion flow onto the magnetic poles.

\item The decrease of $R_\textmd{\scriptsize{bb}}$ toward the lower luminosity was reconfirmed independently based on the coronal optical depth.

\item As to Obs$.$ 5, Obs$.$ 6 and Obs$.$ 7, Model 1 is considered to be more appropriate than Model 2, although the reality could be in between.

\end{enumerate}



\end{document}